\documentclass{appolb}
\usepackage{epsfig}
\def\be{\begin{equation}}
\def\ee{\end{equation}}

\begin{document}
\eqsec 
\title{Total Cross-sections and   Bloch-Nordsieck Gluon  Resummation}

\author{G. PANCHERI
\address{INFN, Frascati National Laboratories, I00044Frascati, Italy}
\and  
R.M. GODBOLE
\address{Centre for High Energy Physics, Indian Institute of Science,
Bangalore,  560 012, India}
\and
A. GRAU
\address{Departamento de F\'\i sica Te\'orica y del Cosmos, 
Universidad de Granada, Spain}
\and 
Y.N. SRIVASTAVA
\address{Physics Department and INFN, University of Perugia, Perugia, 
Italy}}

\maketitle
\begin{abstract}
The physics underlying the fall and eventual rise in various total 
cross-sections at high energies has been investigated over a decade 
using a model based on the Bloch-Nordsieck resummation in QCD. Here a 
brief review of our latest results is presented and comparison made with 
experimental data on $pp$, $\gamma \ 
proton$ and $\gamma \gamma$ 
total cross-sections.
\end{abstract}
  
\section{Introduction}

Total cross-sections at high energies provide significant information
about the distribution of the constituents and the nature of their
interaction at very short distances. Even though QCD is the fundamental
theory for strong interactions, our lack of knowledge about the confinement
of quarks and glue has not allowed a first principle determination of
hadronic total cross-sections and hence one has to resort to phenomenological
models. Over several years we have developed and refined a model based on a 
Bloch Nordsieck(BN) resummation of soft partons. Through it,  we have been 
improving our understanding of the observed variations in 
the cross-sections, in a quantitative way. Details of our work and its
evolution, can be followed through references \cite{tg,corsetti,rohini,grau,
godbole1,jhep,pacetti,godbole2}.  A short summary of data for various 
processes and their comparison with our model predictions are discussed
in the subsequent sections. Given the paucity of space, we shall only focus
on the energy dependences in $pp$, $\gamma p$ and $\gamma \gamma$ reactions
and the uncertainities therein present. 

\section{QCD and the energy dependence of total cross-sections}
In Fig.(1), we show  a comparison of the energy dependence in 
different processes. The data show a clear initial fall and eventual rise 
in the total cross-sections for all processes.
 \begin{figure}[ht] 
     \begin{center}
\epsfig{file=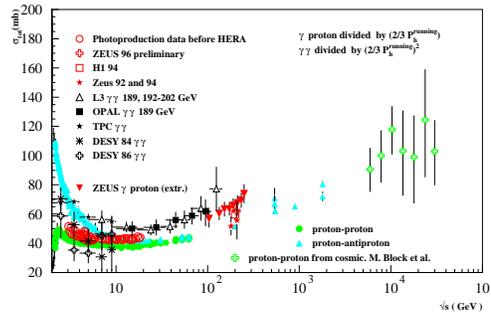,width=7.0cm}
     \end{center}
     \caption{pp/$\bar p p$,$\gamma$p and $\gamma \gamma$ total cross-sections.
 The scaling factor to compare photons on the same scale is obtained from 
quark counting rules and VMD.} 
     \label{fig:allfig1}
     \end{figure}

\par The uncertainties in the data for  $\gamma p$ \cite{HERAZ,HERAH1,DIS,camil} and
 $\gamma \gamma$\cite{L3,OPAL} are shown in Fig.(2) and Fig.(3).
\begin{figure}[ht]
     \begin{center}
\epsfig{file=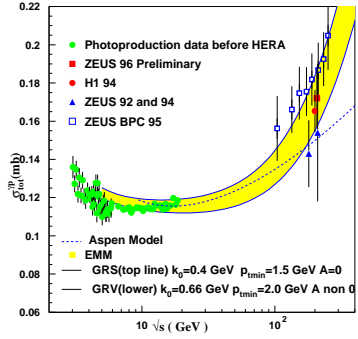,width=5.7cm}
     \end{center}
     \caption{Photoproduction data compared with predictions from the Aspen 
model and the EMM.}
     \label{fig:fig2}
     \end{figure}

 \begin{figure}[ht]
     \begin{center}
     \begin{tabular}{cc}
        \mbox{\epsfig{file=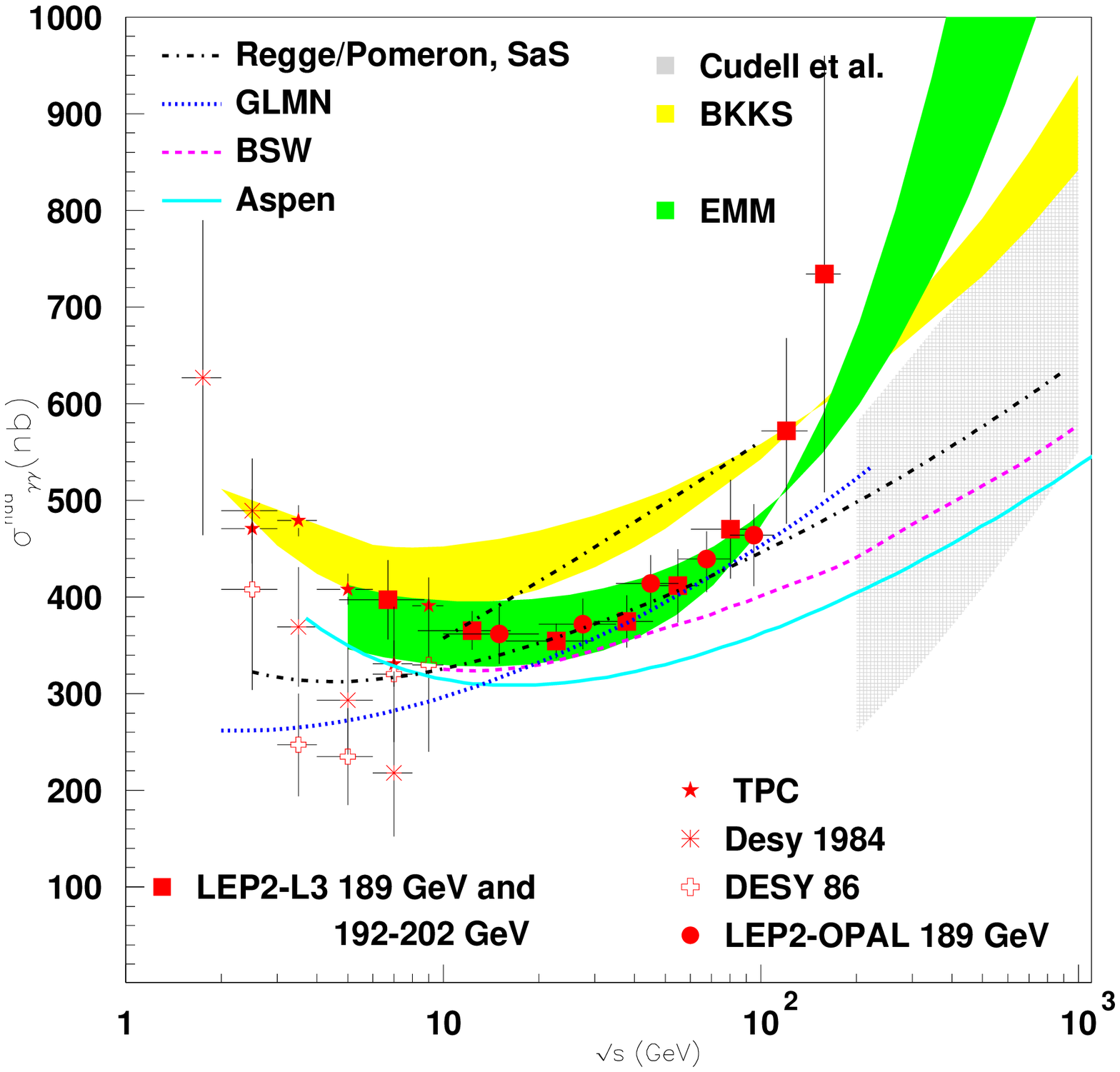,width=5.7cm}}&
        \mbox{\epsfig{file=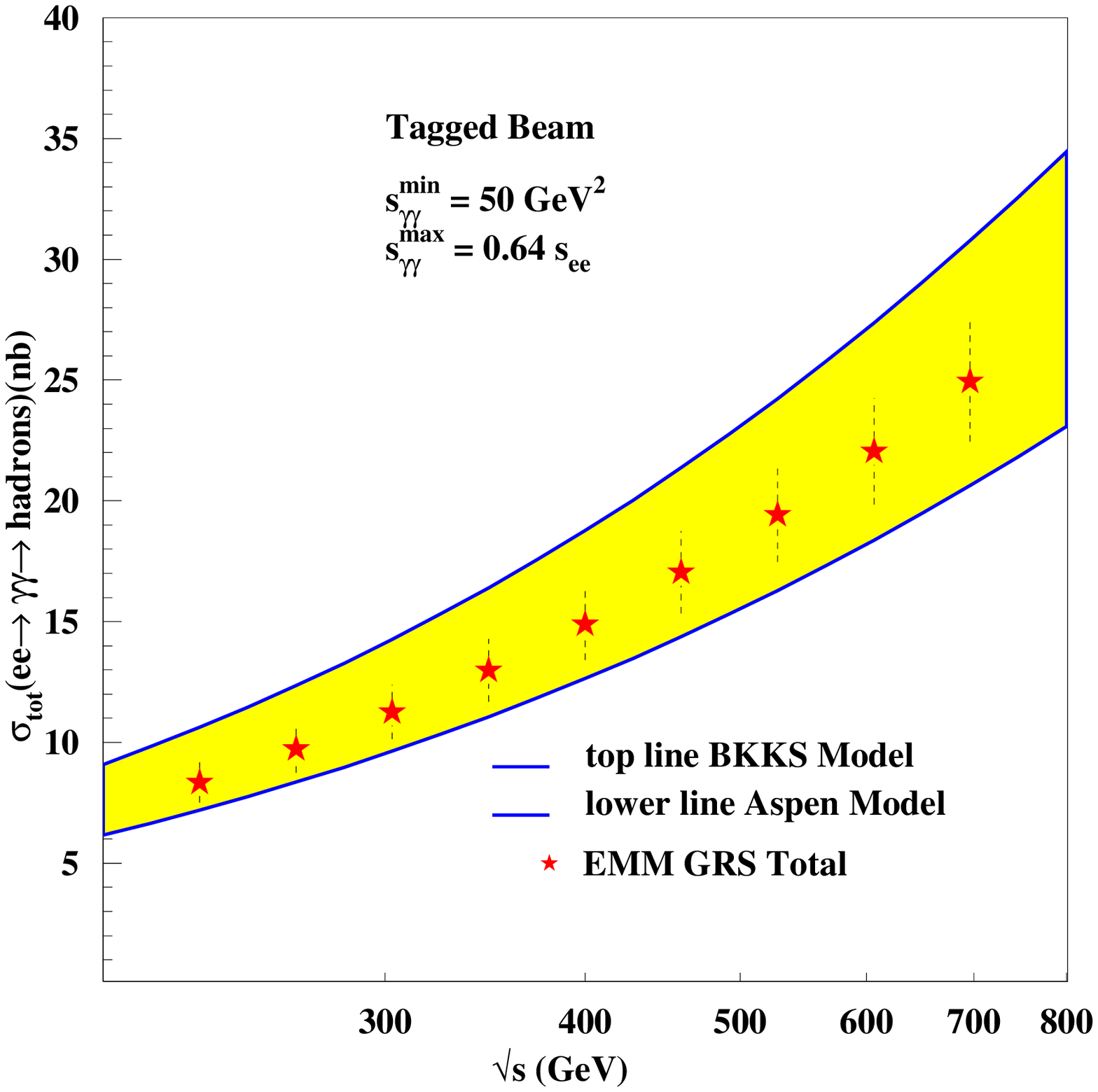,width=5.7cm}}
     \end{tabular}
     \end{center}
     \caption{At left we show $\gamma \gamma$ cross-section data compared with 
predictions from different models. At right the corresponding predictions for 
$e^+e^-$ hadronic cross-sections in the EMM, Aspen and BKKS models.}
     \label{fig:fig3}
     \end{figure}

Theoretically, perturbative QCD provides a natural mechanism to explain the 
rise with energy of total cross-sections. As the hadronic c.m. energy 
increases from $5$ to $10^4 \ GeV$ the number of parton collisions 
increases. In this approach. it is the rise with energy of the jet 
cross-section 
\be
\sigma_{jet}= \int_{p_{tmin}}^{\sqrt s/2} d p_t
\int_{4 p_t^2/s}^1 d x_1 \int_{4 p_t^2/(x_1 s)}^1 d x_2 \sum_{i,j,k,l}
f_{i|a}(x_1) f_{j|b}(x_2) \frac { d \hat{\sigma}_{ij \rightarrow kl}(\hat{s})}
{d p_t},
\ee
which drives the rise of the total cross-section. This quantity depends 
strongly on $p_{tmin}$, the minimum transverse momentum of the produced jets 
and can be calculated by convoluting the parton densities for protons and 
photons. To satisfy unitarity, the jet cross-sections are embedded into 
the eikonal formalism. In this Eikonal Minijet Model (EMM) the total 
cross-section is given by
\be
\sigma_{tot}=2 \int d^2{\vec b}[1-e^{-n(b,s)/2}]
\ee
where $n(b,s)$ is the average number of inelastic 
collisions at impact parameter $b$. 
Introducing a separation between the soft and hard contributions and 
assuming  factorization of the impact parameter and energy dependence 
we can write
\be
n(b,s)=n_{soft}+n_{hard}=A_{soft}(b)\sigma_{soft}+A_{jet}(b)\sigma_{jet}
\ee
where $\sigma_{jet}$ drives the rise and the function $A(b)$ represents the 
impact parameter distribution of partons in the collision.
\par In the simplest EMM formulation $A(b)$ is obtained through convolution 
of the electromagnetic form factors of the colliding particles, $i.e.$
\be
A_{ab}(b)\equiv A(b;k_a,k_b)={{1}\over{(2\pi)^2}}\int d^2{\vec q} 
e^{i q\cdot b} {\cal F}_a(q,k_a){\cal F}_b(q,k_b) 
\ee  
 This model is unable to describe - without further adjustments - the 
experimental data for total cross-sections in the all energy range. This can 
be seen in Fig.(2) where data for $\gamma p$ cross-section are compared with 
a band corresponding to different sets of parameters in the EMM. 
Similarly, we show in Fig.(3) $\gamma \gamma$ cross section data and
its comparison with various models, EMM \cite{rohini}, Regge-Pomeron\cite{SAS}, Aspen\cite{aspen}, BSW\cite{ttwu}, GLMN\cite{glmn}, Cuddell et al.\cite{newfit}, BKKS\cite{BKKS}. The predictions for   $e^+e^-
\rightarrow hadrons$ cross-section appear in Fig. (3) for the Aspen\cite{aspen}, EMM \cite{rohini} and BKKS
\cite{BKKS} models. 

\subsection{Energy dependence of soft gluon emission}

A more realistic EMM is obtained taking into account soft gluon emission 
from initial state valence quarks. In this model, the impact parameter 
distribution of partons is obtained as the Fourier transform of the transverse 
momentum distribution of the colliding partons computed through soft gluon 
resummation techniques. The resulting expression is
\be
\label{abn}
A_{BN}={{e^{-h(b,s)}}\over{\int d^2{\vec b}\ e^{-h(b,s)}}}
\ee
where
\be
\label{hb}
h(b,s)={{8}\over{3\pi}}\int_0^{q_{max}}{{dk}
\over{k}}
\alpha_s(k^2)\ln({{q_{max}+\sqrt{q_{max}^2-k^2}}
\over{q_{max}-\sqrt{q_{max}^2-k^2}}})[1-J_0(kb)]
\ee
The upper limit $q_{max}$ is the maximum energy allowed to each single soft 
gluon emitted in the collision and can be calculated for hard processes 
(those with  $p_t^{parton}\ge p_{tmin}$) by averaging over the valence parton 
densities,$i.e.$
\be 
\label{qmaxav}
M\equiv <q_{max}(s)>={{\sqrt{s}} 
\over{2}}{{ \sum_{i,j}\int {{dx_1}\over{ x_1}}
f_{i/a}(x_1)\int {{dx_2}\over{x_2}}f_{j/b}(x_2)\sqrt{x_1x_2} \int_{z_{min}}^1
 dz (1 - z)}
\over{\sum_{i,j}\int {dx_1\over x_1}
f_{i/a}(x_1)\int {{dx_2}\over{x_2}}f_{j/b}(x_2) \int_{z_{min}}^1 (dz)}}
\ee
with $z_{min}=4p_{tmin}^2/(sx_1x_2)$. As $q_{max}$ depends on the energy 
of the colliding partons, the impact parameter distribution Eq.(\ref{abn})
will be energy dependent. 
The behaviour of $q_{max}$ with energy is shown in Fig.(4) where the upper line
is the one obtained with Eq.(\ref{qmaxav}) and the lower curve are  the 
$q_{max}$ values through which  the soft part $n_{soft}(b,s)$  has been calculated 
phenomenologically to describe $pp$ scattering at low energy.
Using these values of $q_{max}$, in Fig.(4) we show the predictions of 
the model for $pp$ and $\bar p p$ total cross-sections with GRV \cite{GRV} densities. 

 \begin{figure}[ht]
     \begin{center}
     \begin{tabular}{cc}
     \mbox{\epsfig{file=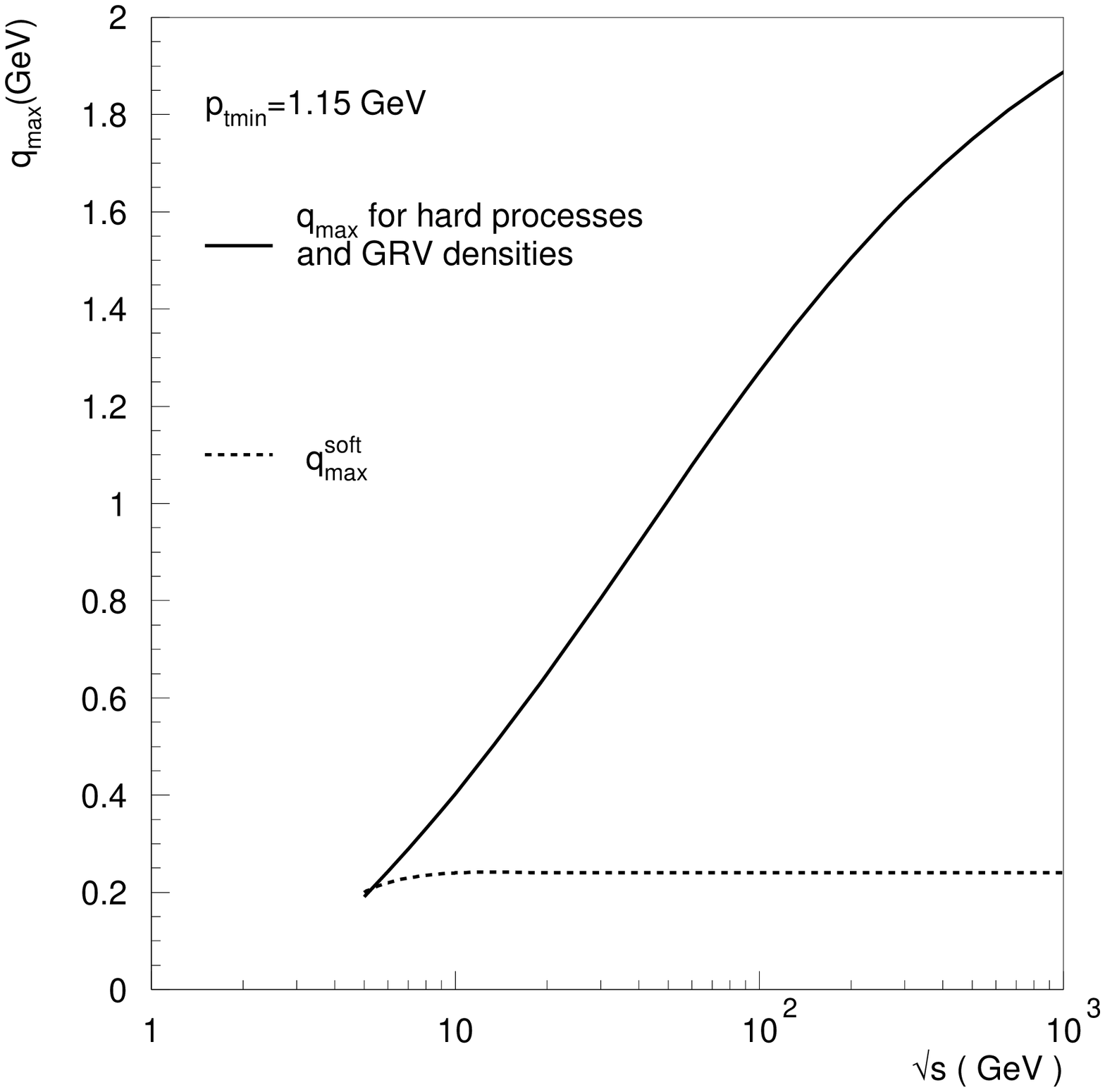,width=5.7cm}}&
     \mbox{\epsfig{file=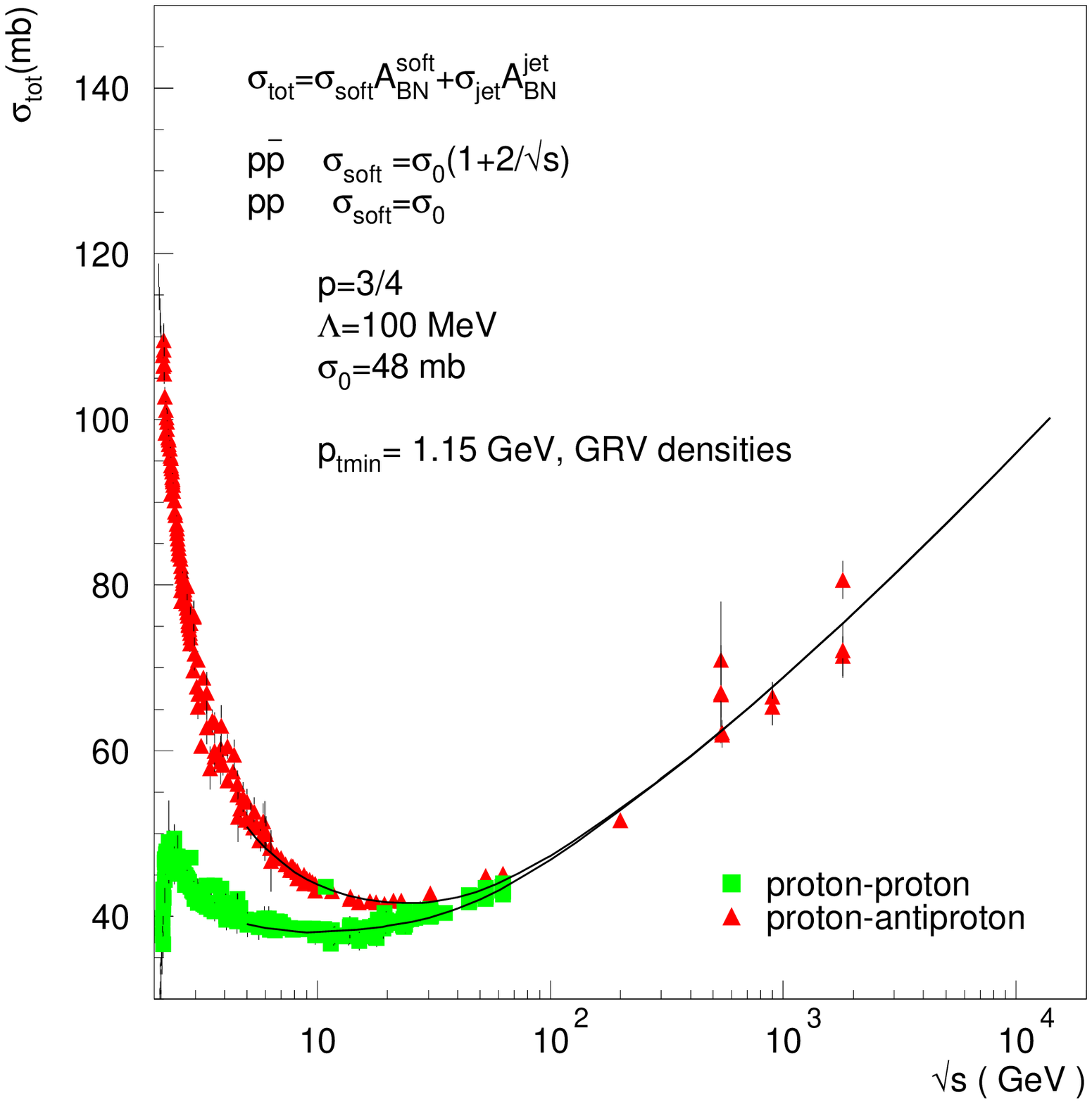,width=5.7cm}}
     \end{tabular}
     \end{center}
     \caption{At left we show the energy dependence of the maximum energy 
allowed to single gluon emission for hard or soft processes. At right we show 
pp and $\bar p p$ total cross-sections data compared with predictions from EMM 
with Bloch-Nordsieck soft gluon resummation.}
     \label{fig:fig4}
     \end{figure}
\par The EMM model  with Bloch-Nordsieck soft gluon resummation have also been 
applied to  $\gamma p$ and $\gamma \gamma$ collisions. Theoretical results 
are compared with experimental data and shown in Fig.(5) for $\gamma p$ and 
in Fig.(6), for $\gamma \gamma$  using two different partonic densities 
for the photon, GRS\cite{GRS} and CJKL\cite{CJKL}. 

 \begin{figure}[ht]
     \begin{center}
     \begin{tabular}{cc}
     \mbox{\epsfig{file=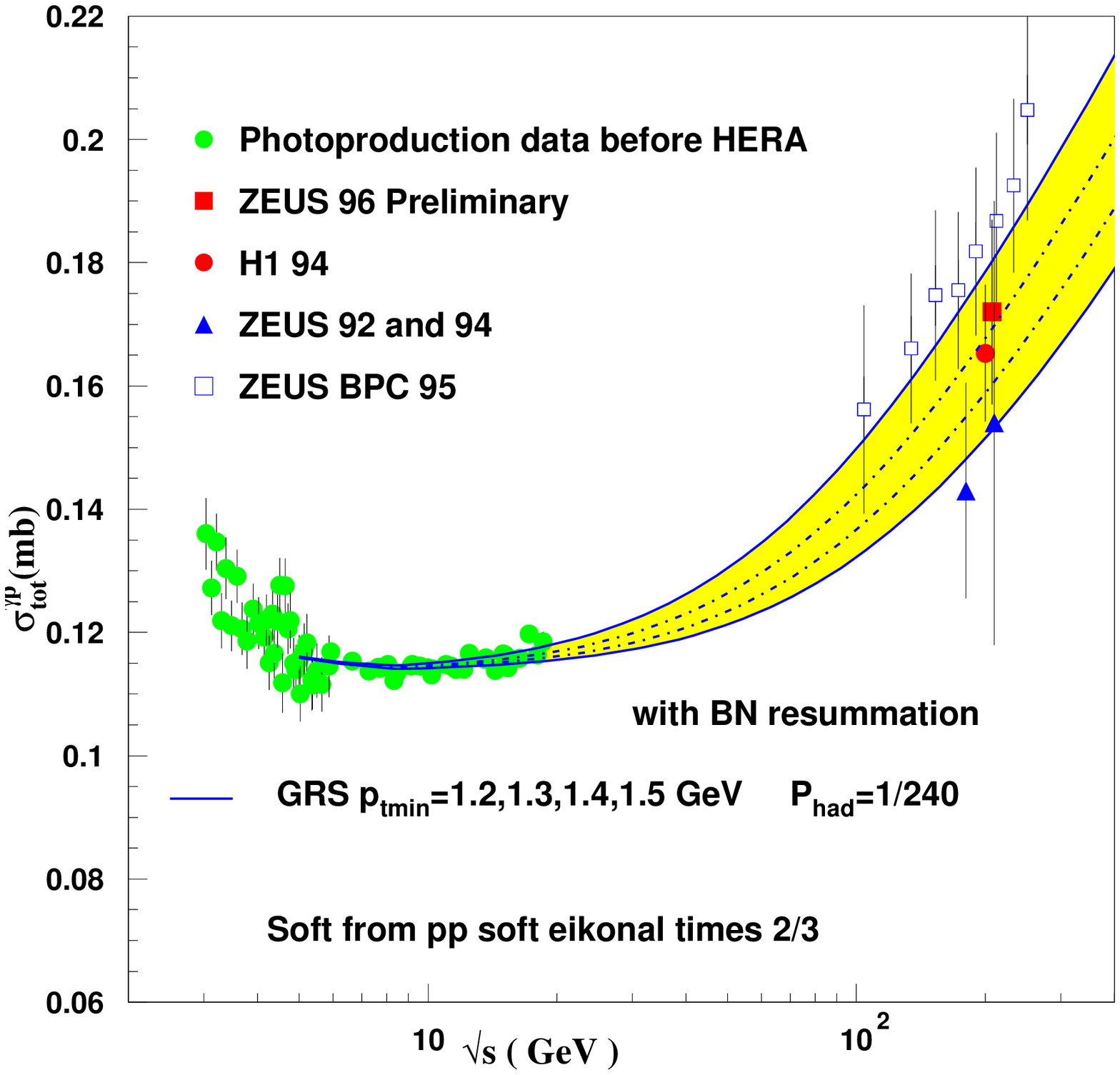,width=5.7cm}}&
     \mbox{\epsfig{file=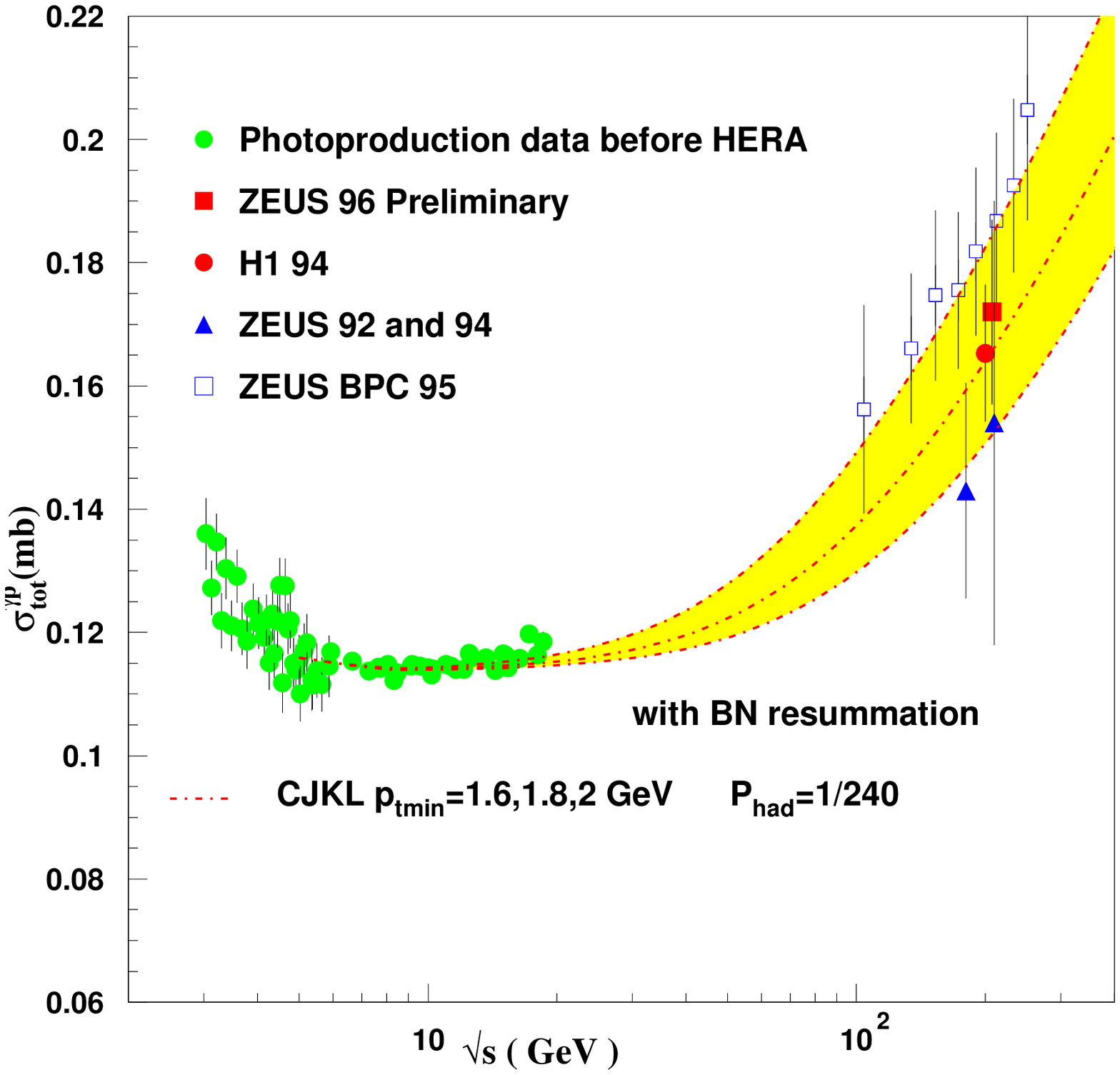,width=5.7cm}}
     \end{tabular}
     \end{center}
     \caption{We show  photoproduction data compared with the soft
     gluon improved EMM for different values of $p_{tmin}$ using GRS densities 
for the photon (at left) and CJKL densities (at right).}
     \label{fig:fig5}
     \end{figure}

 \begin{figure}[ht]
     \begin{center}
     \begin{tabular}{cc}
     \mbox{\epsfig{file=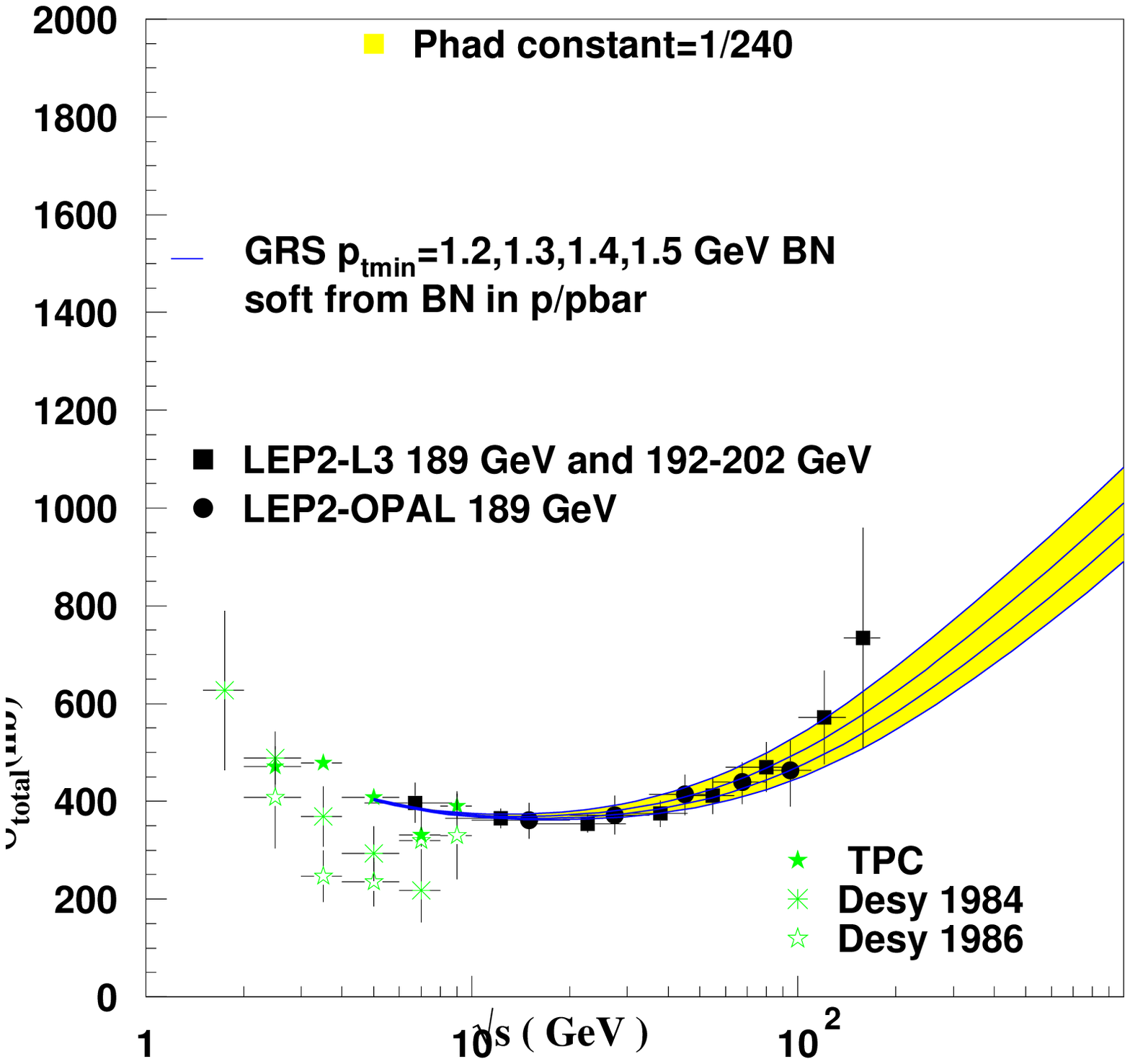,width=5.7cm}}&
     \mbox{\epsfig{file=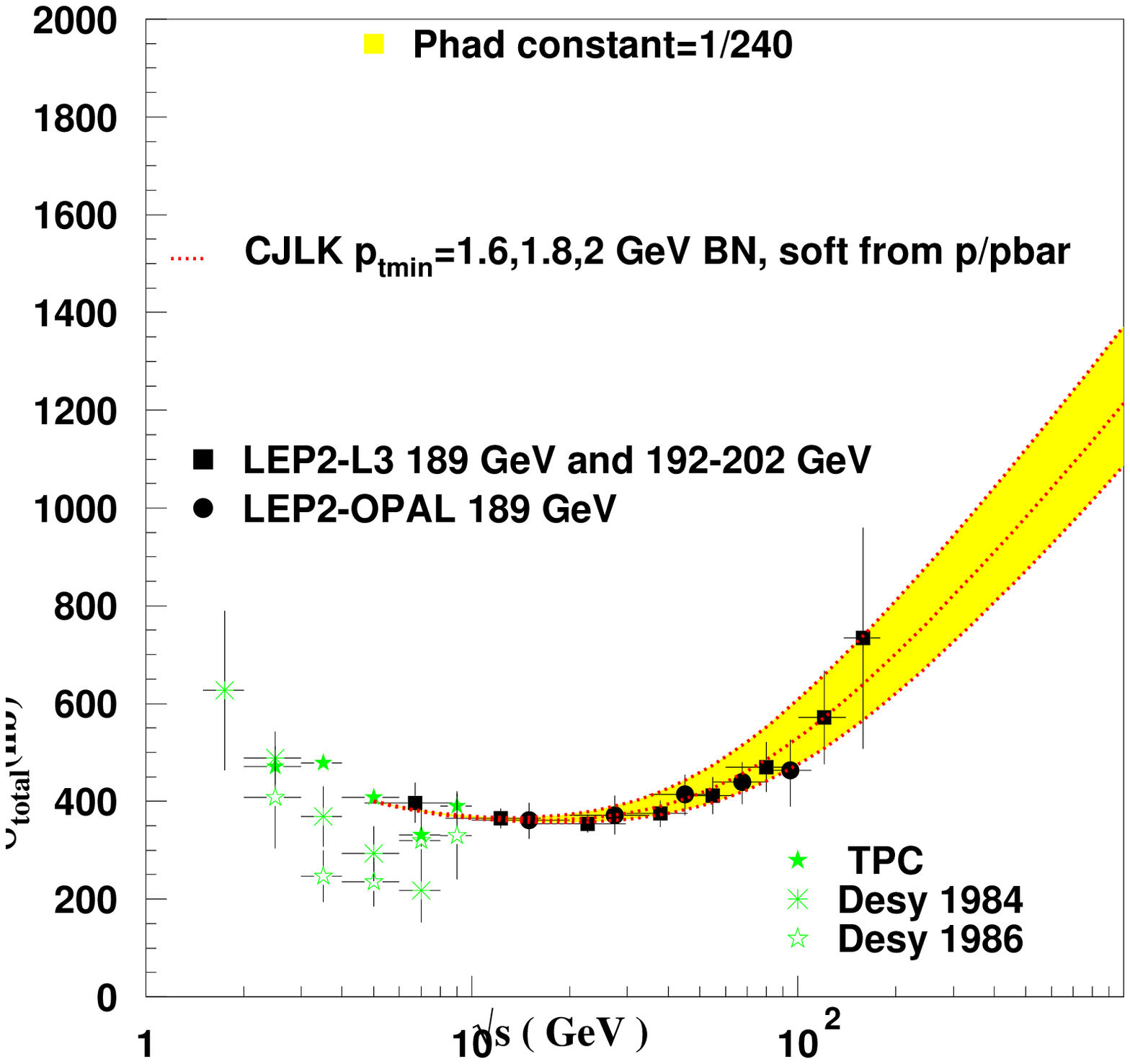,width=5.7cm}}
     \end{tabular}
     \end{center}
     \caption{We show $\gamma \gamma$ cross-section data compared with the soft
     gluon improved EMM using GRS photon densities (at left) and CJKL densities
 (at right).}
     \label{fig:fig6}
     \end{figure}

\section{Conclusion}

In this brief survey, we have presented a comparison of our model predictions
with available data for various processes. The BN resummed gluon distributions
appear to describe quite adequately the rise and fall visible in the data.
Experimentally, there are still significant uncertainites. Theoretically,
we need a better understanding of the $q_{max}$ parameter for the soft
part. 

\section{Acknowledgments}
This work was supported in part by EU Contract CEE-311.
RG wishes to acknowledge the partial support of the Department of
Science and Technology, India, under project number SP/S2/K-01/2000-II.
AG acknowledges support from MCYT under project number FPA2003-09298-c02-01.

\end{document}